\documentstyle[aps,prl,epsfig]{revtex} 
\begin{document}
%\baselineskip = 2\baselineskip
\draft
%\twocolumn[\hsize\textwidth\columnwidth\hsize\csname@twocolumnfalse\endcsname
\title{Scale-dependence of Elastic Constants in the Decoupled
Lamellar Phase of Tethered, Crystalline Membranes }
\author{Yashodhan Hatwalne$^1$\cite{hat} and Sriram
Ramaswamy$^2$\cite{sri}} 
\address{ $^1$Raman Research Institute,
Bangalore - 560 080, INDIA\\ $^2$ Centre for Condensed Matter
Theory, Department of Physics, Indian Institute of Science,
Bangalore - 560 012, INDIA\\} 
\maketitle 
\widetext
\begin{abstract} 
We analyse the effect of thermal
fluctuations on the elastic constants of the decoupled lamellar
phase of tethered, crystalline membranes.  Using a momentum-shell
renormalization group technique, we show that the smectic-A -like
compressional elastic constant, the in-plane Lam\'{e}
coefficients, and the cross-coupling elastic constant vanish as $
(\ln q)^{- a} $ whereas the bend elastic constant diverges as $
(\ln q)^{b} $, with $ a, b > 0$, as the wavenumber $ q $ tends to
zero.  The exponents $ a $ and $ b $ satisfy the relation $ a + 3
b = 2 $. 
\end{abstract}

%\vspace{1 cm} 
%\twocolumn\vskip.5pc]
%\narrowtext 
%\noindent 

In lamellar phases of tethered, crystalline membranes an equilibrium
phase transition from a conventional three-dimensional solid to a
decoupled phase has been predicted.  For large enough mean
separation between neighbouring membranes, the lamellar phase
gets decoupled; the membranes can slide past each other and
rotate relative to each other without any elastic energy cost
~\cite{toner}.  This phase differs from the sliding columnar
phase ~\cite{ohern} which is composed of two-dimensional smectic
layers stacked one on top of other and has a rotation modulus for
relative rotation of the layers.  Lamellar phases of polymerized
membranes are likely candidates to search for the phase
transition from the uniaxial three dimensional solid phase to the
decoupled phase.  Since there is no first-order elastic coupling
between successive membranes, translational order decays
algebraically.  The absence of the shear modulus and the
rotational modulus described above has interesting consequences
for sound propagation and damping in this phase ~\cite{hatwalne}.
\\ In this paper we study the static elastic properties of the
decoupled lamellar phase.  Our principal result is that nonlinear
strains demanded by rotational invariance lead to the
renormalization of the elastic constants; thermal fluctuations do
not destroy the decoupled lamellar phase.  We find that in
addition to the smectic-A -like compressional elastic constant
$B$, and the bend elastic constant $K$, the in-plane Lam\'{e}
coefficients $\lambda$ and $\mu$, and the cross-coupling elastic
constant $\gamma$ become scale-dependent quantities because of
thermal fluctuations.  Asymptotically, $B$, $\lambda$, $\mu $ and
$\gamma $ vanish as $(\ln q)^{-a}$ whereas $K$ diverges as $(\ln
q)^{b}$, $a, b > 0$, for the wavenumber $q \rightarrow 0$, with
$a + 3 b = 2$.  Note that the in-plane Lam\'{e} coefficients
vanish in a manner different from that for a single tethered,
crystalline membrane ~\cite{aronovitz}.  We also point out that
the analysis of this problem in ~\cite{guitter} is erroneous.  \\

To derive these results, we begin by considering the nonlinear
elasticity theory for the decoupled phase ~\cite{hatwalne}.  We
use Cartesian coordinates with the axis of the system along $x$.
The elastic Hamiltonian is 
\begin{equation}
\label{eq:Hamiltonian} 
{\it H} = \int d^{3}x \left[ \frac{B}{2}\
E_{xx}^{2} + \frac{K}{2}\ ( \nabla_{\bot}^{2} u_{x} )^{2} +
\frac{\lambda}{2} \ U_{ii}^{2} + \mu U_{ij}^{2} + \gamma E_{xx}
U_{ii} + \Delta_{1} E_{xx} + \Delta_{2} U_{ii} \right],
\end{equation} 
where the terms with coefficients $\Delta_{1}$ and
$\Delta_{2}$ are the counterterms discussed below.  In
(\ref{eq:Hamiltonian}), 
\begin{equation} 
\label{eq:ustrain}
U_{ij} = \frac{1}{2}\ \left( \nabla_{i}^{\bot} u_{j}^{\bot} +
\nabla_{j}^{\bot} u_{i}^{\bot} - \nabla_{i}^{\bot} u_{l}
\nabla_{j}^{\bot} u_{l} \right), 
\end{equation} 
and
\begin{equation} 
\label{eq:estrain} 
E_{xx} = \nabla_{x} u_{x} -
\frac{1}{2}\ \left( \nabla u_{x} \right)^{2} 
\end{equation}
define the full Eulerian strain tensor, with the superscript $
{\bot} $ representing components in the $yz$-plane.  \\ 
 We now adopt the methods of ~\cite{grinstein} to carry out the
momentum-shell renormalization group analysis of
(\ref{eq:Hamiltonian}).  Power counting shows that the anharmonic
terms in (\ref{eq:Hamiltonian}) are marginal.  We confine the
wavevectors of the Fourier transformed field variables to a
cylindrical region, so that $- \infty \leq q_{x} \leq \infty$ and
$0 \leq \mid{\bf q}_{\bot}\mid \leq \Lambda $, where $ \Lambda$
is of order the inverse lattice constant in the membranes (i.e.,
in the $yz$-plane).  We rescale the field variables
anisotropically:
\begin{eqnarray} 
\label{eq:rescale} 
q_{x} & \rightarrow & b^{2} q_{x}, \nonumber \\ 
{\bf q}_{\bot} & \rightarrow & b {\bf q}_{\bot}, \nonumber \\ 
u_{x}(q_{x}, {\bf q}_{\bot}) & \rightarrow & Z_{x} u_{x}(b^{2} q_{x}, b {\bf q}_{\bot}),
\nonumber \\ 
{\bf u}_{\bot}(q_{x}, {\bf q}_{\bot}) & \rightarrow & Z_{\bot} 
{\bf u}_{\bot}(b^{2} q_{x}, b {\bf q}_{\bot}),
\end{eqnarray} 
where $ b = e^{- \delta l} $.  We integrate out
the degrees of freedom in the thin momentum shell $e^{- \delta l}
\Lambda \leq \mid{\bf q}_{\bot}\mid \leq \Lambda$ through a
one-loop perturbative calculation described below.  \\ 
From the elastic Hamiltonian (\ref{eq:Hamiltonian}), it is apparent that
$u_{x} $ is coupled nonlinearly to $ {\bf u}_{\bot} $ as well as
to itself.  Also, the correlations in $u_{x} $ are more strongly
singular than those in $ {\bf u}_{\bot} $.  The effect of these
singular correlations can be systematically taken into account by
using standard graphical perturbation theory.  Our first step is
to identify the free propagator for the $u_{x} $ -field.  This is
easily done by integrating out the $ \bf{u}_{\bot} $ -field from
the probability distribution for the harmonic theory:
\begin{equation} 
\label{eq:Heff} 
{\it H}_{eff}[u_{x}] = - k_{B} T
\ln \int {\it D} {\bf u}_{\bot} e^{- \beta {\it H}_{0}[u_{x},{\bf
u}_{\bot}]}, 
\end{equation} 
where ${\it H}_{0} $ is the harmonic
part of ${\it H} $ in (\ref{eq:Hamiltonian}), $ k_{B} $ is the
Boltzmann constant, $ T $ is the temperature, and $ \beta =
1/k_{B}T $.  We thus obtain the free propagator 
\begin{equation}
\label{eq:g0}
 G_{0}^{(x)}({\bf k}) = t/\left( k_{x}^{2} +
\lambda_{0}^{2} k_{\bot}^{4} \right), 
\end{equation} 
where $ t = k_{B} T/B_{eff} $, $ B_{eff} = B - [ \gamma^{2} ( \lambda + 2
\mu) /( \lambda + \mu)^{2}] $, and the length $ \lambda_{0} =
(K/B_{eff})^{1/2} $.  Notice that (\ref{eq:Heff}) has the same
form as that of the elastic Hamiltonian for the smectic-A phase.
Integrating out the $ u_{x} $ -field from the probability
distribution gives the free propagator 
\begin{equation}
\label{eq:gperp} 
G_{0}^{(\bot)}({\bf k}) = 
\langle u_{i}^{\bot}({\bf k}) u_{i}^{\bot}({\bf k}) \rangle 
=  \frac{1}{\beta}\ \left[ \lambda + \mu - \frac{\gamma^{2}
k_{x}^{2}}{(Bk_{x}^{2} + K k_{\bot}^{4} ) k_{\bot}^{2} } \
\right] 
\end{equation} 
for the $ {\bf u}_{\bot} $ -field.  Next,
we identify the rescaling factors in (\ref{eq:rescale}) as $
Z_{x} = b^{4} $, and $ Z_{\bot} = b^{4} $.  The flow equations
for this renormalization group are:  
\begin{eqnarray}
\label{eq:flow} 
\frac{dB}{dl}\ & = & - \beta
\frac{(B+\gamma)^{2}}{16 \pi} wt, \\ 
\frac{d \lambda}{dl}\ & = &
- \beta \frac{(\lambda + \mu + \gamma)^{2}}{16 \pi}\ wt, \\
\frac{d \mu}{dl}\ & = & - \beta \frac{\mu^{2}}{4 \pi}\ wt, \\
\frac{d \gamma}{dl}\ & = & \beta \frac{(\lambda + \mu + \gamma)(B
+ \gamma)}{16 \pi}\ wt, \\ 
\frac{dK}{dl}\ & = & \frac{wK}{32
\pi}\ + \frac{Bt}{32 \pi}\ \frac{(\lambda + \mu +
\gamma)^{2}}{(\lambda + \mu)B - \gamma^{2}}\ \left[
\frac{(\lambda_{1}^{2} -
\lambda_{0}^{2})}{\lambda_{0}(\lambda_{2}^{2} -
\lambda_{0}^{2})}\ + \frac{(\lambda_{2}^{2} -
\lambda_{1}^{2})}{\lambda_{2}(\lambda_{2}^{2} -
\lambda_{0}^{2})}\ \right], 
\end{eqnarray} 
where $ w =t/\lambda_{0}^{3} $, $ \lambda_{1}^{2} = K/B $ and $
\lambda_{2}^{2} = (\lambda + \mu)K/[(\lambda + \mu )B -
\gamma^{2} ] $.  These flow equations simplify considerably if $
\gamma(l) \rightarrow 0 $ (which we show below to be the case),
so that $ \lambda_{0} = \lambda_{1} = \lambda_{2} $.  In
particular, the flow equation for $ K $ then reads
\begin{equation} 
\label{eq:Kflow} 
\frac{dK}{dl}\ = \frac{wK}{32
\pi}\ \left[ 1 + \frac{( \lambda + \mu)}{B}\ \right].
\end{equation}
With the simplified flow equation (\ref{eq:Kflow}), the coupled
set of flow equations can be analyzed for the asymptotic 
$ (q \rightarrow 0) $ behaviour of the elastic constants, yielding the
result 
\begin{equation} 
\label{eq:  asymp} 
B, \lambda, \mu,\gamma \asymp ( \ln q)^{- a},K \asymp ( \ln q)^{b},
\end{equation} 
where $ a, b > 0 $, and $ a + 3 b = 2 $.  We note
that $ \gamma $ indeed vanishes as $ q \rightarrow 0 $, which
ensures the consistency of our calculations.  \\
We choose the coefficients $ \Delta_{1} $ and $ \Delta_{2} $ by
demanding that the averages $ \langle \nabla_{x} u_{x} \rangle $
and $ \langle u_{ii} \rangle $ be zero.  From the graphical
structure of the theory we see that this choice of $ \Delta_{1} $
and $ \Delta_{2} $ also cancels terms proportional to $
q_{\bot}^{2} $ generated in the perturbative calculation
~\cite{ward}.  \\

Let us now look at the calculation of ~\cite{guitter}.  In that
work, the invariant 
\begin{equation} 
\label{eq:uxx} U_{xx} =
\nabla_{x} u_{x} - \frac{1}{2}\ (\nabla_{x} u_{x})^{2} -
\frac{1}{2}\ (\nabla_{x} u_{i}^{\bot})^{2} 
\end{equation} 
is used instead of $E_{xx} $ in constructing (\ref{eq:Hamiltonian}). 
It then follows that the vertex 
$ (B/2) (\nabla_{x} u_{x})(\nabla_{x} u_{i}^{\bot})^{2} $ 
gives a correction
\begin{equation} \
\label{eq:dB} 
\Delta B \propto \int dq_{x}
q_{x}^{4} \int d^{2} q_{\bot} 
\langle u^{\bot} u^{\bot} \rangle^{2} 
\propto q_{\bot}^{- 2} \, \forall q_{x}
\end{equation} 
to $B$, since the correlation function 
$ \langle u^{\bot} u^{\bot} \rangle \propto q_{\bot}^{- 2} \, \forall q_{x} $, 
as can be seen from (\ref{eq:gperp}).  Moreover, the same
vertex also generates a term of the form 
$ (\nabla_{x} u^{\bot})^{2} $, 
which drives the system into the three
dimensional uniaxial solid phase, thus destroying the decoupled
lamellar phase itself.  This is because $ \it{H} $ (with $U_{xx} $ 
replacing $ E_{xx} $) does not respect the symmetry of the
decoupled lamellar phase, in that it then includes an elastic
energy cost (at the anharmonic level) for shearing the membranes
past each other.  \\

We thank Tom Lubensky and John Toner for useful discussions.  \\


\begin{references}
\bibitem{ward} This can be verified either by direct evaluation
or by using Ward identities.
\end{references}
\end{document}